\documentclass[
    ,final            % use final for the camera ready runs
  ]
  {aipproc}

\layoutstyle{6x9}

\def \ergsec{\hbox{erg s$^{-1}$}}
\def \kevcmsq{\hbox{keV cm$^{2}$}}

\def\spose#1{\hbox to 0pt{#1\hss}}
\def\ltsim{$\mathrel{\spose{\lower 3pt\hbox{$\sim$}}
        \raise 2.0pt\hbox{$<$}}$\thinspace}
\def\gtsim{$\mathrel{\spose{\lower 3pt\hbox{$\sim$}}
        \raise 2.0pt\hbox{$>$}}$\thinspace}
\def \msun {${\rm M_\odot}$}

\newcommand\solar{\hbox{{$Z_{\odot}$}}}

\newcommand{\chandra }{{\em Chandra}}

\newcommand{\xmm }{{\em XMM}}

\begin{document}

\title[AGN feedback in AWM 4 and NGC 5044]{AGN Feedback in Galaxy Groups: the two interesting cases of AWM 4 and NGC 5044}


\author{Fabio Gastaldello}{
  address={IASF-Milano, INAF, via Bassini 15, Milano 20133, Italy}
  ,altaddress={Department of Physics and Astronomy, University of California at Irvine, 4129 Frederick Reines Hall, Irvine, CA 92697-4575}
}

\author{David A. Buote}{
  address={Department of Physics and Astronomy, University of California at Irvine, 4129 Frederick Reines Hall, Irvine, CA 92697-4575}
}

\author{Fabrizio Brighenti}{
  address={Dipartimento di Astronomia, Universit\`a di Bologna, via Ranzani 1, Bologna 40127, Italy}
  ,altaddress={UCO/Lick Observatory, University of California at Santa Cruz, 1156 High Street, Santa Cruz, CA 95064}
}

\author{William G. Mathews}{
  address={UCO/Lick Observatory, University of California at Santa Cruz, 1156 High Street, Santa Cruz, CA 95064}
}

\author{Pasquale Temi}{
  address={SETI Institute, Mountain View, CA 94043; and Department of physics and Astronomy, University of Western Ontario, London ON N6A, 3K7, Canada}
}

\author{Stefano Ettori}{
  address={INAF, Osservatorio Astronomico di Bologna, via Ranzani 1, Bologna 40127, Italy}
}

\begin{abstract}
We present AGN feedback in the interesting cases of two groups: AWM 4 and 
NGC 5044.
AWM 4 is characterized by a combination of
properties which seems to defy the paradigm for AGN heating in
cluster cores: a flat inner temperature profile indicative of a past, major
heating episode which completely erased the cool core, as testified by the
high central cooling time (> 3 Gyrs) and by the high central entropy
level ($\sim$50 keV cm$^2$), and yet an active central radio galaxy with 
extended radio lobes out to 100 kpc, revealing recent feeding of the central 
massive black hole. A recent Chandra observation has revealed the presence of 
a compact cool corona associated with the BCG, solving the puzzle of the 
apparent lack of low entropy gas surrounding a bright radio source, but 
opening the question of its origin.
NGC 5044 shows in the inner 10 kpc a pair of cavities together with a set of
bright filaments. The cavities are consistent with a recent AGN outburst as 
also indicated by the extent of dust and H${\alpha}$ emission even though the
absence of extended 1.4 GHz emission remains to be explained. The soft X-ray
filaments coincident with H${\alpha}$ and dust emission are cooler than those 
which do not correlate with optical and infrared emission, suggesting that
dust-aided cooling can contribute to the overall cooling. For the first time 
sloshing cold fronts at the scale of a galaxy group have been observed in 
this object.
\end{abstract}

\maketitle

%%%%%%%%%%%%%%%%%%%%%%%%%%%%%%%%%%%%%%%%%%%%
%% MAINMATTER
%%%%%%%%%%%%%%%%%%%%%%%%%%%%%%%%%%%%%%%%%%%%

\section{AGN Feedback in Galaxy Groups}

A growing number of clusters and elliptical galaxies have deep  
multi-wavelength data (X-rays, radio and optical) to study the rich 
phenomenology of cool cores and AGN feedback in a detailed spatially resolved 
fashion, whereas only an handful of groups with such coverage exists and 
therefore, ``unfortunately, AGN heating is not as well studied in groups as 
in clusters'' \citep{McNamara.ea:07}. 
Examination of AGN feedback at the mass scale of groups
is valuable because, although the scale of outbursts in groups is less
energetic and often on a smaller spatial scale than in clusters, the impact can
be even more dramatic than in rich clusters due to
the shallower group potential.
Here we present the cases of the rich phenomenology of 
the cores of two brigh galaxy groups, AWM 4 and NGC 5044.

\section{The puzzle of the core of AWM 4}

AWM 4 is a poor cluster whose X-ray emission is extended and
regular and the peak of the X-ray emission is
coincident with the brightest central galaxy (BCG) NGC 6051. It has a 
bolometric luminosity of $3.93\pm0.06\times10^{43}$ \ergsec\ and a mass 
weighted temperature of $2.48\pm0.06$ keV within 455 kpc, and a mass of 
$1.04\pm0.10 \times 10^{14}$\msun\ within $r_{500} = 708\pm23$ kpc.
It therefore lies at the low end of the mass and 
temperature range defining clusters of galaxies. 28 galaxy members
have been identified, most are absorption line systems with a strong
concentration of early-type galaxies in the center and with a smooth gaussian
velocity distribution centered at the velocity of NGC 6051; 
the velocity dispersion of the system is 
$439^{+93}_{-58}$ km s$^{-1}$ \citep{Koranyi.ea:02}.
NGC 6051 is considerably brighter than the galaxies around it and recent 
deep $R$-band imaging \citep{Zibetti.ea:09} has  
established the fossil nature of this system.
NGC 6051 harbors a powerful radio source with 1.4 GHz flux of $607\pm22$ mJy 
\citep[from the NVSS catalog,][]{Condon.ea:98}.

The \xmm\ observation 
\citep[see the analysis reported in][]{Gastaldello.ea:08*1} revealed a unique 
temperature profile for a relaxed object, with 
an isothermal core out to 200 kpc \citep[as found also in][]{OSullivan.ea:05} 
and then a decline at large radii. They also show a clear abundance gradient, 
from $\sim 0.2$ \solar\ at 400 kpc 
to $1.2\pm0.1$ in the inner 20 kpc, another indication of 
a fairly relaxed system \citep[e.g.,][]{De-Grandi.ea:04}.
To characterize the entropy profile, determined by computing the
adiabatic constant $K = kTn_e^{-2/3}$ we fitted a power law plus a central 
constant $K = K_0 + K_{100}\,(r/100\,\rm{kpc})^{\alpha}$, finding 
$K_0 = 52 \pm 6$ \kevcmsq, $K_{100} = 122 \pm 51$ \kevcmsq\ and 
$\alpha = 1.14\pm0.08$ with $\chi^{2}$/dof = 7/6. The elevated central entropy 
of AWM 4 is reflected in the long 
central cooling time, $3.0\pm0.2$ Gyr in the inner bin. In this regard AWM 4
seems to share the same characteristics of the two radio-quiet clusters, 
A 1650 and A 2244, described in \citet{Donahue.ea:05} and would be consistent 
with the hypothesis that a major past AGN outburst has completely erased the 
cool core. 
But the current 1.4 GHz radio activity implies a recent (not much more than 
$\sim 10^8$ yr ago) feeding 
of the central black hole, which is unlikely to come from the ambient hot gas
which has an order of magnitude longer cooling time. AWM 4 shows a surprising 
lack of low entropy gas surrounding an active and bright radio source 
at 1.4 GHz: its anomalous nature stands out 
clearly in a plot of radio power versus central entropy  $K_0$, as in the plot 
of the BCGS of the \chandra\ sample of
\citet{Cavagnolo.ea:08}: there are very few outliers showing radio emission 
above a threshold of 30 \kevcmsq. These objects may be special cases of BCGs
with embedded coronae, very compsct "mini-cooling cores" (\ltsim 5 kpc) with
low temperatures and high densities; coronae are a low entropy environment and 
may provide the thr conditions necessary for gas cooling to proceed 
\citep{Sun.ea:07}. The \xmm\ data could not rule out the presence of 
low-entropy gas at scales smaller than the inner 20 kpc and indeed a recent 
\chandra\ observation has revealed the presence of a cool ($\sim$ 1 keV) 
compact and extended corona associated with the BCG NGC 6051. 
If this discovery solves
the mistery of the apparent lack of low entropy gas surrounding a bright radio
galaxy and makes AWM 4 consistent with the idea that "every BCG with a strong 
radio AGN has an X-ray cool core", being the cool core 
a large scale bonafide cool core or a 
small corona \citep{Sun:09*1}, it also opens the question of the origin of a 
corona in a relaxed cluster, given the fact that coronae, starting from the 
first objects observed, the BCGs in the Coma cluster 
\citep{Vikhlinin.ea:01*1}, seem to be generally associated with merging 
subclusters.

%%%%%%%%%%%%%%%%%%%%%%%%%%%%%%%%%%%%%%%%%%%%
%% Sample figure:
%%
%% The option [height=...] scales the picture to the given height,
%% without it it would be printed at its nominal size
%%%%%%%%%%%%%%%%%%%%%%%%%%%%%%%%%%%%%%%%%%%%

\begin{figure}
  \includegraphics[height=.3\textheight]{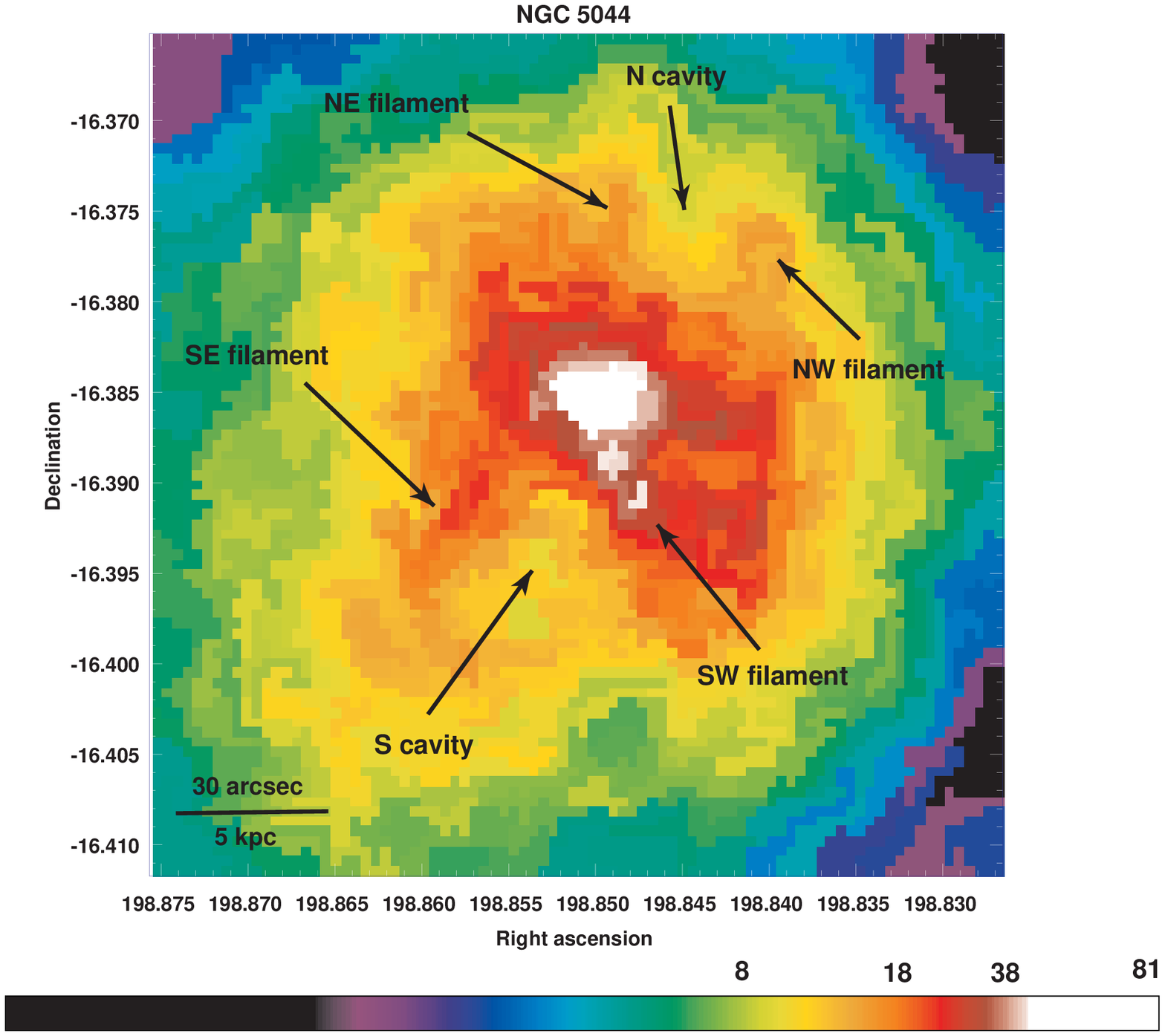}
  \includegraphics[height=.3\textheight]{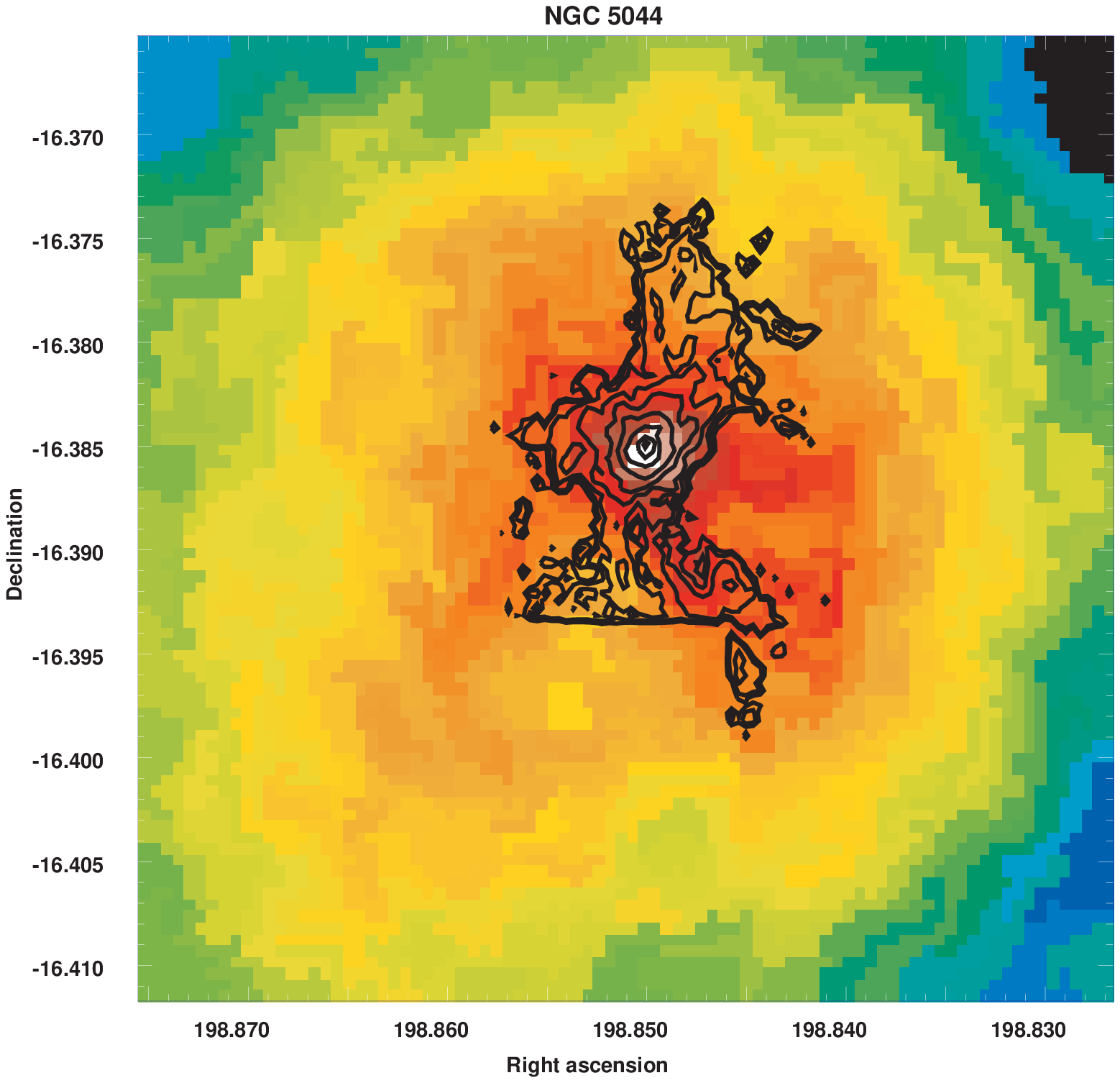}
  \caption{\emph{Left panel:} \chandra\ 0.5-5 keV X-ray image of 
the inner 30x30 kpc of NGC 5044. Interesting features are highlighted.
\emph{Right panel:} Surface brightness contours of 
the H$\alpha$ + [N \tiny$\rm{II}$\footnotesize] map taken from 
\citet{Caon.ea:00} superimposed on the X-ray image.}
\label{fig:1}
\end{figure}

\section{Cavities, Filaments and Cold Fronts in NGC 5044}

The galaxy group NGC 5044 is one of the brightest groups in X-rays: 
our estimate, using the new \chandra\ and \xmm\ data, for the bolometric 
X-ray luminosity within $r_{500}=443 h_{70}^{-1}$ kpc is 
$1.05\pm0.06 \times 10^{43} h_{70}^{-2}$ \ergsec.
An H$\alpha$ nebula is present in the core of NGC 5044 showing an extended
filamentary structure \citep[e.g.,][]{Caon.ea:00}.
NGC 5044 is also remarkable because {\em Spitzer} data show extended 
cold dust emitting at 70 $\mu m$ and extended 8 $\mu m$ excess 
(likely arising from PAH, polycyclic aromatic hydrocarbon, molecules) 
extending out to several kpc and spatially coincident with the H$\alpha$ 
emitting nebulosity and the brightest soft X-ray 
emission \citep{Temi.ea:07*1}. 
As proposed in \citet{Temi.ea:07*1}, current evidence is consistent with an 
internal origin of this dust, which has been buoyantly transported from the 
galactic core out to several kpc into the hot X-ray emitting gas following an 
AGN outburst. Because of its short lifetime ($\sim 10^7$ yrs)
to sputtering destruction by thermal ions, this dust is a spatial tracer 
of extremely transient events.
A two-dimensional analysis of the \chandra\ and \xmm\ data available at that 
moment \citep[see][]{Gastaldello.ea:09} has revealed the wealth of structures
present in the core of this object. The \chandra\ image 
(see Fig.\ref{fig:1}, left panel) shows two depressions 
in surface brightness and multiple filamentary structures, some of them 
connected to the presence of the cavities. The temperature map of the inner
10 kpc reveals that the filament spatially coincident with the H$\alpha$ and 
dust filament is cooler than any other region on the map. Therefore the 
\chandra\ data show the impact of the energy release from the AGN 
on the hot X-ray emitting gas, reinforcing the scenario proposed in 
\citet{Temi.ea:07*1}. However it is  remarkable that a pair of cavities 
close to the nuclear source are lacking extended high frequency radio 
emission. This is contrary to what is generally observed in particular 
in clusters of galaxies. 
They also strengthened the association of dust, 
H$\alpha$ and soft X-ray emission (see Fig.\ref{fig:1}, right panel): the 
presence of the cavity in the North
likely explains the origin of the N H$\alpha$ filament. X-ray filaments are 
present at both sides of the cavities, but only
the ones with the presence of dust are showing optical emission and 
\emph{cooler} X-ray emission. \citet{Temi.ea:07*1} showed that the
cospatiality of these features can be explained as the result of 
dust-assisted cooling in an outflowing plume of hot dusty gas: dust can cool 
buoyant gas to $10^4$ K, which emits the optical emission lines observed.
It is unfortunate that the deep H$\alpha$ observation of \citet{Caon.ea:00}
is affected by a CCD defect in the southern region co-spatial to the X-ray 
cavity; we have obtained IFU observations at the VLT to map in detail the
H$\alpha$ emission in NGC 5044.

At 31 kpc and 67 kpc two surface brightness discontinuities have been 
detected, confirmed to be a pair of cold fronts by the spectral analysis.
The widely accepted scenario for the presence of cold fronts in relaxed 
cool core systems is that they are due to
sloshing of the cool gas in the central gravitational potential, which is
set off by minor mergers/accretions \citep{Ascasibar.ea:06}.
This is also suggested by the peculiar velocity of the brightest 
galaxy NGC 5044 with respect to the mean group velocity \citep{Mendel.ea:08}.
NGC 5044 is the first relaxed group for which
sloshing cold fronts have been discussed in close similarity to the 
ubi\-qui\-tous ones detected in relaxed clusters.

%%%%%%%%%%%%%%%%%%%%%%%%%%%%%%%%%%%%%%%%%%%%%%%%
%% BACKMATTER
%%%%%%%%%%%%%%%%%%%%%%%%%%%%%%%%%%%%%%%%%%%%%%%%

\begin{theacknowledgments}
We thank P.J Humphrey and J. Sanders for the use of their software and N. Caon
for the use of the H$\alpha$ image of Fig.\ref{fig:1}.
\end{theacknowledgments}



\bibliographystyle{aipproc}   % if natbib is available

\bibliography{gasta}

\IfFileExists{\jobname.bbl}{}
 {\typeout{}
  \typeout{******************************************}
  \typeout{** Please run "bibtex \jobname" to optain}
  \typeout{** the bibliography and then re-run LaTeX}
  \typeout{** twice to fix the references!}
  \typeout{******************************************}
  \typeout{}
 }

\end{document}